\documentclass[acmsmall]{acmart}
\AtBeginDocument{%
  \providecommand\BibTeX{{%
    Bib\TeX}}}

\setcopyright{acmlicensed}
\copyrightyear{2026}
\acmYear{2026}
\acmISBN{978-1-4503-XXXX-X/2018/06}

\begin{document}

\title{Digital Agriculture Sandbox for Collaborative Research}

\author{Osama Zafar}
\affiliation{%
  \department{Dept. of Computer and Data Sciences}
  \institution{Case Western Reserve University}
  \city{Cleveland}
  \state{Ohio}
  \postcode{44106}
  \country{USA}
}
\email{oxz23@case.edu}

\author{Rosemarie Santa González}
\affiliation{%
  \department{School of Computer Science}
  \institution{Georgia Institute of Technology}
  \city{Atlanta}
  \state{Georgia}
  \postcode{30332}
  \country{USA}
}
\email{rosemarie.santa@gatech.edu}

\author{Alfonso Morales}
\affiliation{%
  \department{Dept. of Planning and Landscape Architecture}
  \institution{University of Wisconsin–Madison}
  \city{Madison}
  \state{Wisconsin}
  \postcode{53706}
  \country{USA}
}
\email{morales1@wisc.edu}

\author{Erman Ayday}
\affiliation{%
  \department{Dept. of Computer and Data Sciences}
  \institution{Case Western Reserve University}
  \city{Cleveland}
  \state{Ohio}
  \postcode{44106}
  \country{USA}
}
\email{exa208@case.edu}

\renewcommand{\shortauthors}{Zafar et al.}

\begin{abstract}
  Digital agriculture generates valuable data that can support research and improve farming practices, but privacy concerns limit data sharing among farmers and researchers. This paper presents the Digital Agriculture Sandbox, a privacy-preserving science gateway that enables collaborative agricultural data analysis through secure workflows. The platform integrates federated learning, differential privacy, and privacy-aware data processing to support similar farmer identification, communication, collaborative model training, and model sharing while maintaining control over sensitive farm data. Unlike our previous work, which introduced the underlying privacy framework, this work focuses on implementing an accessible gateway that operationalizes these techniques through user-facing services. The containerized platform provides identity management, secure data management, workflow execution, and model interaction capabilities, enabling agricultural stakeholders to collaborate without requiring extensive computational expertise.
\end{abstract}




\received{20 February 2007}
\received[revised]{12 March 2009}
\received[accepted]{5 June 2009}

\maketitle

\section{Introduction}
Agriculture has been the cornerstone of human society and its evolution for thousands of years. Over the years, it evolved from basic hunting and gathering to large-scale planting and harvesting. In the era of digital innovation and data-driven evolution, digital agriculture is playing a vital role in driving productivity, aiding research, and evolving the entire field of agriculture. The large-scale use of electronic devices and machines in the agricultural sector has enabled the collection of massive amounts of data from fields and their use in research and development. Collected data is used to develop next-generation seeds, highly effective fertilizers, insecticides, and pesticides, as well as to train machine learning models for irrigation management, disease detection, field monitoring, yield prediction, and various other research initiatives, thereby transforming cultivation and harvest practices \cite{pretty2008agricultural, tilman2002agricultural}. However, due to the sensitive nature of data collected from farms, farmers are hesitant to share it with others \cite{WAKWEYA2023100698, WESTERMANN2018283}. The data can be easily leveraged against farmers, resulting in adverse consequences such as insurance rate increases, unfavorable pricing, discrimination, and resource manipulation \cite{demets, wiseman2019farmers, taylor2018climate}. This lack of access to farmers hinders research and development in the field. To unlock the full potential of agricultural data, access to small-scale farm data is paramount; thus, it is imperative to develop solutions that allow researchers to access the data they need in a privacy-preserving manner that protects and safeguards the privacy and livelihoods of farmers.  

In this work, we present the Digital Agriculture Sandbox, a privacy-aware science gateway designed to operationalize collaborative agricultural data analysis. The goal of this work is to provide an accessible platform where farmers and researchers can collaborate while maintaining control over sensitive agricultural data. Unlike our previous work, which introduced the underlying privacy-preserving framework and similarity-based collaboration approach, this paper focuses on the implementation of a usable gateway that integrates these techniques into practical workflows. The main contributions of this work are: (1) the design and implementation of a containerized science gateway for privacy-preserving agricultural collaboration; (2) user-facing workflows for dataset management, similar farmer discovery, communication, collaborative model training, and model sharing; and (3) deployment of privacy-aware computational services that enable machine learning workflows without requiring farmers to manage complex infrastructure.

The remainder of this paper is organized as follows. Section~\ref{background} provides technical background on the privacy-preserving technologies used in the sandbox. Section~\ref{framework} describes the Digital Agriculture Sandbox framework, its key functionalities, the deployment architecture, implementation details, and practical applications. Finally, Section~\ref{future} and \ref{conclusion} provide future directions and conclusions.

\section{Technical Background}
\label{background}
In this section, we define the specific terms being used in the remainder of the paper. Federated learning enables multiple entities to collaboratively train machine learning models without exchanging raw data. Each participant trains a local model and shares updates for aggregation into a global model. Membership inference attacks aim to determine whether a specific sample was included in model training, motivating privacy-preserving mechanisms such as differential privacy. Model inference refers to using trained models for predictions; in black-box settings, users only access model outputs without internal model information.

\section{Digital Agriculture Framework}
\label{framework}

While our prior work \cite{zafar2025} focused on developing privacy-preserving mechanisms for secure agricultural collaboration, this work transforms those mechanisms into an accessible science gateway. The sandbox integrates the privacy framework with user management, communication tools, workflow automation, and model-sharing capabilities. It enables farmers and researchers to upload data to their accounts and perform collaborative operations through privacy-aware workflows.

\subsection{Sandbox's underlying architecture}

Sandbox is based on the privacy-preserving data-sharing and collaborative-research architecture proposed in \cite{zafar2025}. The proposed architecture combines techniques such as Principal Component Analysis \cite{Pearson}, local differential privacy \cite{Duchi, Kairouz}, Laplacian noise addition \cite{Dwork2013}, and federated learning to allow farmers to perform operations on distributed data in a privacy-preserving manner. Individual farmers upload farm data to their accounts, which is not shared with other farmers. However, computations such as identifying other farmers based on data similarities and training machine learning models on distributed data can be performed in a privacy-preserving manner. For more details on the architecture and privacy guarantees it provides, please refer to \cite{zafar2025}.

\subsection{Key features}
\textit{\textbf{Identity management}} functionality enables users to access their accounts on the platform, upload and manage datasets, perform computational operations, and train models. We use the Texas Advanced Computing Center (TACC) as the identity provider. Users can create accounts through supported identity providers and use the same credentials to access the sandbox. The \textit{\textbf{Upload}} functionality allows farmers to securely upload and manage their datasets, with access restricted to the account owner. To ensure compatibility with commonly used agricultural systems, uploaded files must be in CSV format with column headers and a target column named ``label''. CSV was selected because many IoT sensors, farm management systems, and agricultural devices already export data in this widely used format, reducing the technical burden on farmers.


The \textit{\textbf{Identify similar farmers}}  functionality enables farmers to discover similar peers for knowledge sharing and collaboration while also identifying suitable participants for privacy-preserving collaborative model training and personalized model development without sharing any raw farmer data using techniques introduced by \cite{zafar2025}. The farmer (the initiator of the operation) selects a dataset from their list of uploaded datasets. Sandbox identifies other farmers (potential collaborators) based on the similarity between the initiator's dataset and their datasets. To ensure uniformity across similarity calculations, farmers with datasets having the same metadata as the initiator's dataset are considered for the similarity calculations. After calculating, an ordered list of farmer profiles (only usernames, no personal identifiers) is displayed in decreasing order of similarity.

\textit{\textbf{Chat}} functionality enables farmers to communicate with "identifying similar farmers" on the platform, facilitating collaboration and knowledge sharing. Farmers can share information with each other based on their comfort level. 

\textit{\textbf{Collaborative model training}} functionality enables farmers to train machine learning models using data from other farmers in a privacy-preserving manner. This functionality leverages federated learning to enhance data privacy and security, as it allows models to be trained without sharing or centralizing sensitive data. To optimize the training process and reduce the number of participating collaborators (farmers), we utilize the "Identify Similar Farmers" functionality and select highly similar farmers to contribute to the model training process. This optimizes the training process by the number of collaborators without significant loss of model utility \cite{zafar2025}.To submit a training request, follow these steps. 
\begin{itemize}
   \item Selecting reference dataset from uploaded datasets.
   \item Entering the name for the model.
   \item Selecting model type from the list of models available.
   \item Selecting model visibility, either public or private. 
   \item Adding how-to instructions for the users. 
   \item Selecting collaborators for training from the list of similar farmers. 
\end{itemize}

To reduce the technical burden on users with limited computational expertise, commonly used configuration options are provided with default values. Users can execute workflows without requiring detailed knowledge of model architectures or optimization parameters.


All submitted model training jobs are executed in the background. Further discussion on model training can be found in Section~\ref{sec:deployment}.

\textit{\textbf{Model Repository}} functionality allows farmers to view available machine learning models. All models have a "model visibility" value set to either "public" or "private". Model visibility controls access for all platform users, including researchers. Public models are available according to platform permissions, while private models remain accessible only to the owner and explicitly authorized collaborators. All models have the following intractable feature available through the platform: 
\begin{itemize}
   \item The information component is accessible by clicking on the info icon, which displays the model's general information, including: model Name, model type, metadata, model owner, model visibility, number of classes, class name, and training status.
   \item Risk quantification component is accessible by clicking on the red shield icon, which displays model inference and risk analysis information such as:
       \begin{itemize}
                \item Model logs show the number of inference queries made by each user and their inference risk score.
                \item Model activity shows individual inference requests with information about the sender, time, and number of queries.
                \item Risk analysis provides the overall risk score for the model, along with a plot illustrating the risk scoring for all users interacting with the model.
        \end{itemize}
        The displayed risk score represents an estimated membership-inference risk associated with model exposure. It is derived from the membership inference analysis done in \cite{zafar2025} and is intended as a risk-awareness indicator rather than a universal security guarantee. The underlying methodology assumes the attacker model and evaluation settings described in \cite{zafar2025}.
   \item The Playground component is accessible by clicking on the play icon, which allows farmers to interact with the model by submitting their data and receiving the model's output. It provides general information on how to submit data for model inference, along with the 'how-to-use' instructions from the model owner, which can be downloaded by clicking the download icon in the top right corner of the card. 
\end{itemize}

\subsection{Deployment}
\label{sec:deployment}

The sandbox follows a distributed computation model where each farmer is associated with a farmer server responsible for processing that farmer’s data. The farmer server acts as the computational node for local operations, including data transformations, privacy-preserving processing, and local model training. Depending on deployment requirements, the farmer server can be hosted locally or on trusted computational infrastructure. In the current deployment, computational resources are provided through TACC infrastructure. The application server coordinates workflows but does not directly access farmer datasets. Requests requiring access to farmer data are executed through the corresponding farmer server. This separation provides logical isolation between users and prevents direct access to one farmer’s data by other users.

\begin{figure}[]
    \centering
    \includegraphics[width=0.5\linewidth]{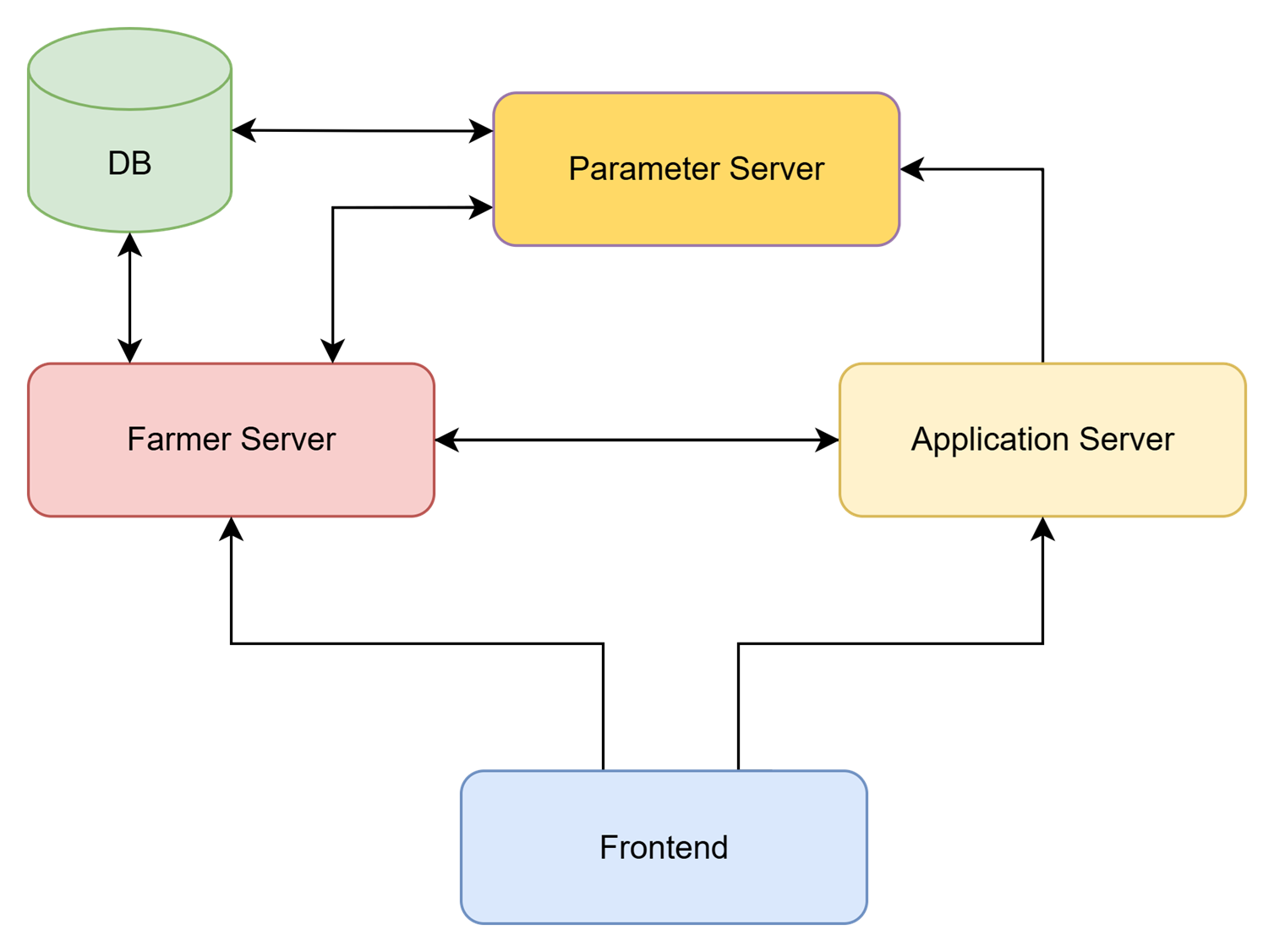} 
    \caption{Illustrates the deployment of the sandbox within a containerized environment and the interaction among the containers.}
    \label{fig:deployment}
    \vspace{-6mm}
\end{figure}

The sandbox environment is a Docker-containerized application comprising five containers: frontend, database, application server, parameter server, and farmer server. Figure~\ref{fig:deployment} shows how deployed containers interact with each other. In the following sections, we provide detailed descriptions of individual containers.
\begin{itemize}
    \item Frontend container serves the web pages and allows the farmers to interact with the platform. 
    \item A database container is used to host a MongoDB server, which stores databases that house uploaded datasets, models, chats, logs, and other information.
    \item Application server container hosts a Flask server, which serves as the core of the application, responsible for all authentication, request orchestration, and computations. It does not have direct access to any of the farmer data or models. It sends requests to the parameter server or farmer server to perform computations on the user's raw data.
    \item Farmer server, executes computations on the farmer's behalf. It loads farmers' data, executes their local computations on their behalf, and deletes all information after the computations are done. It digitally segregates information from different farmers and keeps all information distributed as if it were physically distributed. This provides a privacy guarantee that only farmers have access to their data, and their data is never stored or processed with that of other farmers. Upon request, the framer server is responsible for training local machine learning models on individual farmers' data, sharing updates to the parameter server, and also computing membership inference risk for trained models.
    \item Parameter server is used to execute model training jobs. Jobs are submitted to the parameter server by the application server, which then orchestrates the training of local models for all collaborators. With the farmer server, it receives model updates and aggregates them to save the final model weights. It also computes risk analysis for a trained model by interacting with the farmer server.
\end{itemize}

We chose our stack purely on technical merit. Docker containerization lets us deploy seamlessly across major cloud providers (AWS, GCP, Azure) or local servers without dependency headaches. We used MongoDB for flexible unstructured data storage, though it can be swapped out if needed. Finally, end-to-end sandbox testing has already verified everything from authentication and dataset management to model training and inference. The platform source code and deployment documentation are available through the project repository~\footnote{Github Repository \url{https://github.com/ICICLE-ai/food-security-sandbox}} \footnote{Deployed Platform \url{https://icicleai.tapis.io/v3/oauth2/idp}}, while a formal usability evaluation with farmers and agricultural stakeholders remains part of future work.

\subsection{Use Cases}
The sandbox platform enables farmers to share sensitive data, addressing their privacy concerns securely. It allows farmers to discover and connect with like-minded individuals, fostering collaboration while maintaining their data privacy. This promotes effective sharing, learning, and development of best practices. For researchers and stakeholders, the platform provides access to collaborative computational workflows, including model training, model sharing, and inference capabilities using authorized datasets and models. Direct researcher data-sharing workflows remain an area of future expansion.

\section{Future Work}
\label{future}
Future work will focus on extending the sandbox capabilities and improving its adoption. We plan to support additional model training options, customization, and hyperparameter tuning, while strengthening safeguards against potential misuse of similarity identification and communication features. We also aim to develop privacy-preserving data-sharing workflows that enable controlled collaboration between farmers and researchers through techniques such as anonymization and differential privacy. Beyond technical improvements, future efforts will investigate usability challenges, farmer onboarding, stakeholder requirements, and adoption barriers for privacy-preserving agricultural data collaboration.

\section{Conclusion}
\label{conclusion}
Digital agriculture has significant potential to improve food production and research, but privacy concerns limit the sharing of valuable farm data. The Digital Agriculture Sandbox addresses this challenge by providing a privacy-preserving science gateway for collaborative agricultural data analysis. By integrating federated learning, differential privacy, PCA, and secure workflow management, the platform enables privacy-aware similarity identification, collaborative model training, and model interaction. Through features such as identity management, secure data handling, communication, and containerized deployment, the sandbox provides an accessible foundation for agricultural collaboration while maintaining control over sensitive farm information.

\begin{acks}
This work was supported by ICICLE: Intelligent CI with Computational Learning in the Environment, funded by the National Science Foundation (NSF) under award number OAC-2112606. The authors also gratefully acknowledge the Texas Advanced Computing Center (TACC) for providing computational resources, infrastructure, and technical support that contributed to the research results reported in this work.
\end{acks}



\bibliographystyle{ACM-Reference-Format}
\bibliography{sample-base}

\appendix






\end{document}